\documentstyle[aps]{revtex}
\begin{document}

\title{A Perturbative Method to solve fourth-order Gravity Field Equations}

\author{
M.Campanelli \and C.O.Lousto\footnote[4]{Present Address: Universidad
Aut\'onoma de Barcelona, IFAE, Grupo de F\'\i sica Te\'orica,
E-08193 Bellaterra (Barcelona), Spain. E-mail: lousto@ifae.es}\footnote[2]
{Permanent Address: Instituto de Astronom\'\i a y F\'\i sica
del Espacio, Casilla de Correo 67 - Sucursal 28, 1428 Buenos Aires,
Argentina. E-mail: lousto@iafe.edu.ar} and J.Audretsch
}

\address{
  Fakult\"at f\"ur Physik der
  Universit\"at Konstanz,
  Postfach 5560 M 674,
  D - 78434 Konstanz, Germany
}

\maketitle

\abstract
We develop a method for solving the field equations of a quadratic
gravitational theory coupled to matter. The quadratic terms are written
as a function of the matter stress tensor and its derivatives in such a way
to have, order by order, a set of Einstein field equations with an effective
$T_{\mu\nu}$. We study the cosmological scenario recovering the de Sitter
exact solution, and the first order (in the coupling constants $\alpha$ and
$\beta$ appearing in the gravitational Lagrangian)
solution to the gauge cosmic string metric and the charged black
hole. For this last solution we discuss the consequences on the
thermodynamics of black holes, and in particular, the entropy -
area relation which gets additional terms to the usual ${1\over4} A$ value.
\endabstract

\pacs ~PACS numbers: 04.60.+n

\section{Introduction}
Higher derivatives gravitational theories have been proposed at classical
level as an extension of Einstein's theory in an attempt to unify other
fields with gravity\cite{1} and to avoid the cosmological singularity
\cite{2}. Quadratic curvature counterterms also appear in the renormalization
of the one-loop semiclassical approximation\cite{4}. The pure Quadratic theory
is renormalizable\cite{5} and asymptotically free\cite{6} although it has
problems with unitarity\cite{5}. Higher order gravitational theories can
lead naturally to inflation \cite{3,7} and arise as the low energy limit
of string theory\cite{8}.

For the sake of definiteness we deal with the following Lagrangian formulation
of quadratic theories
\begin{eqnarray}
S=S_{G}+S_{m}=&&\int d^4x\sqrt{-g}\bigg\{-2\Lambda + R + \alpha R^2\cr
&&+\beta R_{\mu\nu}R^{\mu\nu} + 2k {\cal L}_m\bigg\}~, \label{1.1}
\end{eqnarray}
where we have dropped the $ R_{\mu\nu\lambda\rho}^2$ term by use of
the Gauss-Bonnet invariant in four dimensions\cite{C62}.

The field equations derived by extremizing the action $S$ are given by
(we use the sign conventions of Ref. \cite{MTW})
\begin{eqnarray}
R_{\mu\nu}-&&{1\over 2}Rg_{\mu\nu}+\Lambda g_{\mu\nu}+\alpha H_{\mu\nu}+
\beta I_{\mu\nu}\cr
&&=kT_{\mu\nu} \doteq -{2k\over\sqrt{-g}}{\partial S_m
\over\partial g^{\mu\nu}} ~,  \label{1.2}
\end{eqnarray}
where we have chosen units such that $c=1$ and $k=16\pi G$, and where
\begin{equation}
H_{\mu\nu}=-2R_{;\mu\nu}+2g_{\mu\nu}\Box R-{1\over 2}g_{\mu\nu}R^2
+2R R_{\mu\nu}~, \label{1.2'}
\end{equation}
and
\begin{eqnarray}
I_{\mu\nu}=&&-2R_{(\mu~;\nu)\alpha}^{\alpha} +\Box R_{\mu\nu}+
{1\over 2}g_{\mu\nu}\Box R\cr
&&+2R_{\mu}^{~\alpha}R_{\alpha\nu}
-{1\over 2}g_{\mu\nu}R_{\alpha\beta}R^{\alpha\beta}~. \label{1.2''}
\end{eqnarray}
Note that the trace of  equation (\ref{1.2}) takes the simple form
\begin{equation}
2(3\alpha + \beta)\Box R -R +4\Lambda=kT ~. \label{1.3}
\end{equation}
In section 2 of this paper we develop a method to find a metric solution of
the classical field equations of higher order gravitational theories with
sources by successive perturbations around a solution to Einstein Gravity,
which will represent for us the zeroth order. In section 3 we apply this method
to find first order solutions in the coupling constants $\alpha$ and $\beta$
for the straight gauge cosmic string and the charged black hole. For this
latter solution the associated thermodynamics is studied and the corrections
to the Bekenstein - Hawking temperature and entropy are discussed.
We end the paper with some further discussion on this first order solutions.

\section{Perturbative Solution}

{}From the Eq.\ (\ref{1.3})
we can see that $R$ satisfies a massive scalar wave equation.
For this field to have a real mass we impose
\begin{equation}
3\alpha + \beta\geq 0~~~~,~~~~  \beta\leq 0 \label{2.1}
\end{equation}
where the last inequality comes from linearizing Eq.\ (\ref{1.2})
and asking a real
mass for the massive spin-two field $\psi_{\mu\nu}$ related to $R_{\mu\nu}$
(see Ref. \cite{AEL93}). Conditions (\ref{2.1}) are called the no-tachyon
constraints.

The coupling constants $\alpha$ and $\beta$ have to be at most of the atomic
scale, since otherwise they could have observable effects in, for instance,
the solar system or binary pulsars. In the  Ref. \cite{MMS86} it was found
that $\alpha<10^{15}l_{pl}^2$ by requiring that the inflationary period be of
sufficient duration (where $l_{pl}$ is of the order of $10^{-33}$ cm). A
similar
bound for $\beta$ can thus be obtained.

We will then consider only small curvatures in our method, such that
\begin{equation}
\alpha |R|\ll 1 ~~~~,~~~~|\beta R_{\mu\nu}|\ll 1~. \label{2.2}
\end{equation}
Thus, we can obtain perturbative solutions to the field
equations in a power series of $\alpha$ and $\beta$. We will take as
the starting  metric, $g_{\mu\nu}^{(0)}$, a solution of Einstein equations
and then, systematically, successive orders of the coupling constant
$\alpha$ and $\beta$.

To {\it zeroth order}
\begin{equation}
R_{\mu\nu}(g_{\mu\nu}^{(0)})-{1\over 2}R^{(0)}g^{(0)}_{\mu\nu}+\Lambda
g^{(0)}_{\mu\nu}=kT_{\mu\nu}(g_{\mu\nu}^{(0)})~, \label{2.3}
\end{equation}
with trace
\begin{equation}
R^{(0)}=4\Lambda-kT^{(0)}
\end{equation}
Once we solve for the Einstein metric $g_{\mu\nu}^{(0)}$, we pursue the
iteration to the {\it first order}
\begin{eqnarray}
R_{\mu\nu}^{(1)}&&-{1\over 2}R^{(1)}g^{(1)}_{\mu\nu}+\Lambda
g^{(1)}_{\mu\nu}\cr
\cr
&&=kT_{\mu\nu}(g_{\mu\nu}^{(1)})-\alpha H_{\mu\nu}(g_{\mu\nu}^{(0)})-
\beta I_{\mu\nu}(g_{\mu\nu}^{(0)})~, \label{2.4}
\end{eqnarray}
where to first order in $\alpha$ and $\beta$ it is enough to consider
in Eq.\ (\ref{2.4}) $H_{\mu\nu}^{(0)}$ and $I_{\mu\nu}^{(0)}$.

By use of Eqs.\ (\ref{2.3}) the right hand side of Eq.\ (\ref{2.4})
can be written as
\begin{eqnarray}
&&T_{\mu\nu}^{(1)eff}\cr\cr
&&=T_{\mu\nu}^{(1)}-{\alpha\over k} H_{\mu\nu}
(g_{\mu\nu}^{(0)}, T_{\mu\nu}^{(0)})-{\beta\over k} I_{\mu\nu}
(g_{\mu\nu}^{(0)}, T_{\mu\nu}^{(0)})~,\label{2.5}
\end{eqnarray}
where
\begin{eqnarray}
{1\over k}H_{\mu\nu}^{(0)}&&\doteq{1\over k}H_{\mu\nu}(g_{\mu\nu}^{(0)},
T_{\mu\nu}^{(0)})=
2T_{;\mu\nu}^{(0)}-2g_{\mu\nu}^{(0)}\Box T^{(0)}\cr\cr
&&+{k\over 2}g_{\mu\nu}^{(0)}
(T^{(0)})^2-2kT^{(0)}T_{\mu\nu}^{(0)}+8\Lambda T^{(0)}_{\mu\nu}
-2\Lambda T^{(0)}g_{\mu\nu}~,\label{2.6}
\end{eqnarray}
and
\begin{eqnarray}
{1\over k}I_{\mu\nu}^{(0)}&&\doteq {1\over k}I_{\mu\nu}(g_{\mu\nu}^{(0)},
T_{\mu\nu}^{(0)})\cr\cr
&&=T_{;\mu\nu}^{(0)} -2T_{~~~~(\mu;\nu)\alpha}^{(0)\alpha}+
\Box T_{\mu\nu}^{(0)}-g_{\mu\nu}^{(0)}\Box T^{(0)}\cr\cr
&&+4\Lambda T_{\mu\nu}^{(0)}-\Lambda T^{(0)}g_{\mu\nu}+{k\over2}
g_{\mu\nu}^{(0)}(T^{(0)})^2\cr\cr
&&-2kT^{(0)}T_{\mu\nu}^{(0)}+2kT^{(0)\alpha}_{~~~~\mu}
T_{\alpha\nu}^{(0)}-
{k\over 2}g_{\mu\nu}^{(0)}T^{(0)\alpha\beta}
T_{\alpha\beta}^{(0)}~. \label{2.7}
\end{eqnarray}
Thus $T_{\mu\nu}^{(1)eff}$ plays the role of an effective energy momentum
tensor in an Einsteinian field equation for $g_{\mu\nu}^{(1)}$.
It is easy to see that $T_{\mu\nu ;}^{{\rm eff}~\nu~}=0$, i.e. satisfies a
conservation law (with respect to $g_{\mu\nu}^{(1)}$).
$T_{\mu\nu}^{\rm eff}$ also inherits
the symmetry properties of $T_{\mu\nu}$. This properties will hold to every
order of the approximation.

We can now generalize to the {\it n-th order} approximation
\begin{eqnarray}
R_{\mu\nu}^{(n)}&&-{1\over 2}R^{(n)}g^{(n)}_{\mu\nu}+\Lambda
g^{(n)}_{\mu\nu}=kT_{\mu\nu}^{{\rm eff}(n)} =
kT_{\mu\nu}(g_{\mu\nu}^{(n)})\cr\cr
&&-\alpha H_{\mu\nu}(g_{\mu\nu}^{(n-1)}, T_{\mu\nu}^{(n-1)})
-\beta I_{\mu\nu}(g_{\mu\nu}^{(n-1)}, T_{\mu\nu}^{(n-1)})
{}~. \label{2.8}
\end{eqnarray}
Expressions (\ref{2.5})-(\ref{2.7})
greatly simplify when $T_{\mu\nu}$ is diagonal and
its components depend essentially on only one coordinate; let
us call it
$r$. This will be the case in the applications we will deal with in
the next section (we will also take $\Lambda=0$ for simplicity).
Thus, in this case
\begin{eqnarray}
{1\over k}H_{\mu\mu}^{(0)}=&&2\bigg\{T_{,rr}\delta^r_{\mu}-\Gamma^r_{\mu\mu}
T_{,r}-g_{\mu\mu}\bigg[g^{rr}T_{,rr}-\cr
&&g^{\alpha\alpha}\Gamma^r_{\alpha\alpha}T_{,r}
-{k\over 4}T^2 +k T T_{\mu}^{\mu}\bigg]\bigg\}~, \label{2.9}
\end{eqnarray}
and
\begin{eqnarray}
{1\over k}I_{\mu\mu}^{(0)}&&=T_{,rr}\delta^r_{\mu}-\Gamma^r_{\mu\mu}T_{,r}\cr
&&-2\bigg[\big(T_{r~,rr}^{r}+\Gamma^{\alpha}_{r\alpha}(T_{r~,r}^{r}-
T_{\alpha~,r}^{\alpha})\big)\delta^r_{\mu}-\Gamma^r_{\mu\mu}T_{\mu ~,r}^{\mu}
\bigg]\cr
&&-g_{\mu\mu}\bigg[2kT T_\mu^\mu-g^{rr}T_{\mu~,rr}^{~\mu}+ g^{\alpha\alpha}
\Gamma^r_{\alpha\alpha}(T_{\mu ~,r}^{\mu}-T_{,r})\cr
&&+g^{rr}T_{,rr}-{k\over 2}T^2-2k(T_{\mu}^{~\mu})^2
+{k\over 2}T^{\alpha\alpha}T_{\alpha\alpha}\bigg] ~.
\label{2.10}
\end{eqnarray}
where the metric dependence in this expressions is with respect to the zeroth
order, i.e. solution of usual Einstein equations, Eq.\
(\ref{2.3}). In Eqs.\ (\ref{2.9}) and
(\ref{2.10}) sum is over $\alpha$ index but not over $\mu$.

\section{First order solutions}

We are now ready to compute the different metrics solution of the fourth order
field equations. Three main astrophysical scenarios where gravity
plays an important role can be studied: cosmology, topological defects
and black holes.

\subsection{De Sitter Universe}

It is a solution of vacuum Einstein equations with cosmological constant.
This solution is useful for describing an inflationary phase in the very
early universe where higher order gravity terms are presumably worth
taking into account.

In our approximation, for $T_{\mu\nu}=0$, we have that de Sitter metric
is a solution of the fourth order field equations order by order (see
Eq.\ (\ref{2.8})), without restrictions for the value of $\Lambda$. Thus, it is
an {\it exact} solution as can be directly verified from the field
equations \cite{BO83}. In fact, in general, vacuum solutions to Einstein
equations (even with cosmological constant), are solutions to the
quadratic theory (the converse, in general, is not true).

The Robertson - Walker metric can be also replaced in Eqs. (\ref{2.4})
and (\ref{2.5})
to find the corrections to the general relativistic results.
However, this approximation breaks down near the cosmological singularity
and we already dispose of some exact solutions (see Ref. \cite{CH90}).

\subsection{Gauge Cosmic Strings}

Our approximation is specially suitable for dealing with the gravitational
field of topological defects since while its effects are of relevance
they are relatively small compared to to the planck scale, thus a first
order approximation in $\alpha$ and $\beta$ should provide acceptable results.

Topological  defects are expected to be formed during phase transitions
driven by the evolving (and cooling down) early universe whenever the
manifold of equivalent vacua after the spontaneous symmetry breaking
is not shrinkable to a point. We can model this process by studying an
$n$-internal components scalar field with a Mexican-hat like effective
potential with an associated energy scale of symmetry breaking, $\eta^2$,
of order $10^{-6}$ for the GUT scale \cite{K76} and coupling constant
$\lambda$. Depending on the dimension
of the vacua manifold the topological defects can be domain walls, cosmic
strings or monopoles and depending on whether the (internal) symmetry
that breaks down is a local (or gauge) or a global one, the formed
topological defects will be localized in a tiny core of size
$r_c\sim {1\over\eta\sqrt{\lambda}}$ or infinitely extended\cite{V85}.

To see explicitly the effects of this higher derivative theory of gravity
we chose one particular topological defect. Since in the next section
we will deal with a spherically symmetric system (a charged black hole)
we will deal here with the local (or gauge) straight cosmic string that
posses cylindrical symmetry. The other topological defects can
be treated in an analogous way (see Ref. \cite{AEL93} for a thorough
account of them).

A straight, static, cylindrically symmetric local string lying along
the $z-axis$ can be characterized by the following energy-momentum tensor
(for $r\gg r_c$)
\begin{equation}
T_{t}^{~t}=T_{z}^{~z}=-{\eta^2\delta(r)\over 2\pi r\sqrt{B(r)}}
{}~~~;~~T_{r}^{~r}=-T_{\theta}^{~\theta}=0~~.\label{3.1}
\end{equation}
For a generic metric of the form
\begin{equation}
ds^2=A(r)(-dt^2+dz^2) + dr^2+r^2B(r)d\theta^2 ~, \label{3.2}
\end{equation}
the exact metric, solution of Einstein equations is given by\cite{V81}.
\begin{equation}
A(r)=1~~,~~~B(r)=B_0=\left(1-k{\eta^2\over 4}\right)^2~~.\label{3.2'}
\end{equation}

By plugging the generic metric (\ref{3.2}) into Eqs.\ (\ref{2.4}) and
(\ref{2.5}) we can
now write the solution to our first order Einstein equation with
effective source $T^{\rm eff}_{\mu\nu}$
\begin{equation}
A(r)=c_1+k\int^r{r''dr''\int^{r''}{{dr'\over r'}\left[T^{{\rm eff}~\theta}
_\theta -T^{{\rm eff}~r}_r\right]}}~,\label{3.4}
\end{equation}
and
\begin{eqnarray}
B(r)=&&c_2-{2kB_0\over r}\int^r dr''\Bigg\{\int^{r''}dr'r'
\bigg[T^{{\rm eff}~\theta}
_\theta\cr
&&-{1\over2}T-(3\alpha+\beta)\Box T\bigg] +A(r'')\Bigg\}~,
\label{3.5}
\end{eqnarray}
where $c_1$ and $c_2$ are arbitrary constants to be conveniently chosen
in such a way to recover general relativistic results.

Explicitly replacing here the form of the gauge cosmic string stress
tensor given by Eq.\ (\ref{3.1}) and upon integration and regularization
of delta squared terms, we obtain the metric
components up to linear terms in $\alpha$ and $\beta$
\begin{equation}
A(r)=1+(2\alpha+\beta)kT~,
\end{equation}
\begin{eqnarray}
B(r)=\left(1-k{\eta^2\over 4}\right)^2&&-4\alpha k\left(1-k{\eta^2\over
2}\right)^2T\cr\cr
&&-{\left(\alpha+{\beta\over 2}\right)k^2\eta^4\over (2\pi)^4}
{1\over r^2}~.\label{3.6}
\end{eqnarray}
We observe here that the corrections to the general relativistic
metric due to including
quadratic terms in the curvature in the gravitational lagrangian
can be classified in two types. The terms in $A(r)$ and $B(r)$ proportional
to the trace of the matter stress tensor, $T$, give localized contributions.
They are different from zero only  in the core of the gauge string and
vanish for $r>r_c$. This is essentially what was found in Ref.  \cite{AEL93},
where we restricted the analysis to linearized terms in the curvature and
only considered its higher derivatives. It is precisely the additional terms
not considered in Ref. \cite{AEL93},
i. e. quadratic in the stress tensor, that generate the non-localized term,
proportional to $r^{-2}$ appearing in $B(r)$. The structure of this term
is such that it is linear in the coupling constants $\alpha$ and
$\beta$ (due to our approximation),
and if one considers they have an associated radius $r_1$, the dependence
$(r_1/r)^2$ is the only extended possible one not divergent as $r\to\infty$.
On the other hand the factor $k^2\eta^4$ already appears in processes such
as particle production by the formation of Cosmic String \cite{P87} and
Global Monopoles\cite{L91}. We see that due to the quadratic gravity terms,
space outside a straight local cosmic string is no longer flat as in
General Relativity, but curved with curvature terms typically going as
$r^{-4}$.
This dependence also appears when one considers the renormalized
energy-momentum tensor due to vacuum polarization\cite{H87,FS87,ML91}.

\subsection{Charged Black Holes}

We shall now study spherically symmetric solutions to the quadratic field
equations in the first order approximation Eqs.\ (\ref{2.4})-(\ref{2.5}), which
represent charged black holes; starting from the General Relativistic
solution, i.e. the Reissner-Nordstr\"om metric,
\begin{equation}
ds^2=g_{tt}(r)dt^2+g_{rr}(r)dr^2+r^2d\Omega^2~,    \label{3.19}
\end{equation}
where $-g_{tt}(r)=g_{rr}(r)^{-1} =(1-2M/r+Q^2/r^2)$ and
$d\Omega^2=d\vartheta^2+\sin^2\vartheta d\varphi^2$.

The non-vanishing components of the energy-momentum tensor are\cite{MTW}
\begin{equation}
T_t^t=T_r^r=-T_{\vartheta}^{\vartheta}=-T_{\varphi}^{\varphi}=
-{Q^2\over r^4}~.      \label{3.20}
\end{equation}
The exact metric, solution of the Einstein equations with source, can be
written as \cite{LS88} (in the Schwarzschild gauge, Eq.\ (\ref{3.19}))
\begin{equation}
g^{-1}_{rr}(r)=1-{2M\over r}+{1\over r}\int_{\infty}^r
{\tilde r^2 T^{{\rm eff}~t}_t d\tilde r}~,   \label{3.21}
\end{equation}
\begin{equation}
g_{tt}(r)=-g_{rr}(r)^{-1} \exp{\left\{\int_{\infty}^r{(T_r^r-T_t^t)^{\rm eff}
\tilde r g_{rr}(\tilde r) d\tilde r}\right\}}~.   \label{3.21'}
\end{equation}
By use of Eq.\ (\ref{3.20}) to compute the $T_{\mu\nu}^{\rm eff}$ (Eq.\
(\ref{2.5})) we find the first order corrections to the energy-momentum tensor
\begin{eqnarray}
\Delta T_r^{{\rm eff}~r}&&={1\over3}\Delta T_t^{{\rm eff}~t}=-{1\over2}\Delta
T_\vartheta^{{\rm eff}~\vartheta}=-{1\over2}\Delta T_\varphi^{{\rm eff}~
\varphi}\cr
\cr
&&={4\beta Q^2\over kr^6}\big(1-{2M\over r}+{Q^2\over r^2}\big)~,
\end{eqnarray}
and replacing it into Eqs.\ (\ref{3.21}) -\ (\ref{3.21'}) we obtain
\begin{equation}
g_{rr}(r)^{-1} \simeq 1-{2M\over r}+{Q^2\over r^2}-{12\beta Q^2\over r^4}
\left({1\over 3}-{M\over 2r}+{Q^2\over 5r^2}\right)~,
\end{equation}
\begin{eqnarray}
-g_{tt}(r) \simeq &&g_{rr}(r)^{-1} e^{2\beta Q^2\over r^4}\cr
&&\cr
\simeq &&1-{2M\over r}
+{Q^2\over r^2}-{2\beta Q^2\over r^4}\left( 1 -{ M\over r}+{Q^2\over 5r^2}
\right)~.\label{3.22}
\end{eqnarray}
We observe here that the $\alpha$ coupling constant does not appear.
This is due  to the fact that the trace of the electromagnetic energy
momentum is zero. In fact, Whitt\cite{W84} has shown that for the quadratic
theories coupled only to the $\alpha$-term  (i.e $\beta=0$ in Eq.\
(\ref{1.2})) there
exist a ``no hair" theorem stating that the only black hole solution
(with spherical symmetry),
must be the Reissner - Nordstr\"om family.
This can be directly seen from our method, since
order-by-order the $\alpha$-contributions vanish, thus leaving us with
the Reissner - Nordstr\"om solution. This is not
the case, of course, when one allows the
$\beta$ terms be different from zero, and expressions Eq.\
(\ref{3.22}) represent
the ${\cal O}(\beta)$ generalized Reissner - Nordstr\"om metric.

Direct inspection of metric (\ref{3.22}) allow us to see the improved result
that generates the perturbative method presented in this paper with respect
to the ``linearized" approach of Ref. \cite{EL93}. There, only terms up to
${\cal O}(r^{-4})$ have been considered.

The validity of this metric will be assured if the condition Eq.\ (\ref{2.2})
holds. In our case, this takes the form $-\beta/r^2\ll 1$. Thus,
Eq.\ (\ref{3.22})
will be a good approximation to the charged black hole solutions
in quadratic theories if
\begin{equation}
r_H\gg\sqrt{-\beta}~, \label{3.23}
\end{equation}
where $r_H$ is the radial coordinate of the event horizon (in
Schwarzschild's gauge).

The geodesic motion acquires an additional small repulsive term in
the effective potential since for the no-tachyon constraint $\beta<0$.
This provides, in principle, a way of detecting the physical effects produced
by the higher order corrections to the gravitational lagrangian.

The radial coordinate of the horizon can be computed directly from
making vanish $g_{tt}(r_H)$ given by Eq.\ (\ref{3.22}),
\begin{equation}
r_H=r_+ +\beta {Q^2\over r^3_+}{\left(1-{3Q^2\over 5r^2_+}\right)\over
\left(1-{Q^2\over r^2_+}\right)}~~,~~ r_+=M+\sqrt{M^2-Q^2}~.
\label{3.25}
\end{equation}
We observe that the quadratic black hole shrinks with respect to the
corresponding general relativistic one.

The extreme black hole will be now reached with a maximal charge lower than the
general relativistic one ($\beta<0$)
\begin{equation}
Q^2_{max}=M^2+{2\over 5}\beta ~,     \label{3.25'}
\end{equation}
thus $r_H$ given by Eq.\ (\ref{3.25}), will remain always bounded.
In fact, $r^{min}_H=M+{\beta\over5M}$ .

The horizon area, $A_H$, will then be
\begin{equation}
A_H=4\pi r^2_+\left[1+{2\beta Q^2\over r^4_+}{\left(1-{3Q^2\over5r^2_+}\right)
\over\left(1-{Q^2\over r^2_+}\right)}\right]~,\label{3.26}
\end{equation}
which is smaller than in the general relativistic case.

The electric potential on the horizon can be independently computed
by use of Maxwell equations in  curved spacetime Eq.\ (\ref{3.22})
\begin{eqnarray}
\Phi_H(r_+)=&&\int_{r_H}^{\infty}{{Qdr\over r^2\sqrt{-g_{tt}g_{rr}}}}\cr
&&\cr
=&&{Q\over r_+}\left[1-{2\beta Q^2\over 5r^4_+}\left({3-2Q^2/r^2_+\over
1-Q^2/r^2_+}\right)\right]~. \label{3.27}
\end{eqnarray}
Let us now turn to the thermodynamical properties of black holes and see
how   they are modified by the quadratic corrections.

We can easily compute the Bekenstein-Hawking temperature from the surface
gravity, $\kappa$:
\begin{equation}
T_H={\kappa\over 2\pi}=-{1\over 4\pi}{g'_{tt}\over\sqrt{-g_{tt}g_{rr}}}
\biggr\vert _{r=r_H}~.   \label{3.24}
\end{equation}

Thus,
\begin{equation}
T_H={1\over4\pi r_+}\left(1-{Q^2\over r^2_+}\right)+{\beta Q^4\over 5\pi r^7_+}
\left({2-Q^2/r^2_+\over 1-Q^2/r^2_+}\right) ~.\label{3.28}
\end{equation}
Since $\beta<0$, we observe that the effect of the quadratic gravitational
theory corrections will be that of decreasing the black hole radiation
temperature with respect to the general relativistic value (with the same
$M$ and $Q$). Leaving thus out open the possibility of switching off black
hole evaporation and leaving behind a charged remnant with a mass of the
order of the Planck mass.

By inverting Eq.\ (\ref{3.26}) for $M$ as a function of $A_H$ and $Q$. We
obtain
the fundamental relation of black hole thermodynamics. Differentiation
of this equation produces the first law,
\begin{equation}
dM={\partial M\over \partial A}\biggr\vert _Q dA+{\partial M\over \partial Q}
\biggr\vert _A dQ=T_HdS+\Phi_HdQ~, \label{3.29}
\end{equation}
that can be used to obtain the entropy of the quadratic black hole, $S$.

Since,
\begin{eqnarray}
{\partial M\over \partial A}\biggr\vert _Q&&={1\over 16\pi r_+}
\left(1-{Q^2\over r^2_+}\right)\cr
&&\times\left[1+{2\beta Q^2\over r^4_+}
{\left(1-{6\over 5}{Q^2\over r_+^2}+{3\over 5}{Q^4\over r_+^4}\right)
\over\left(1-{Q^2\over r_+^2}\right)^2}\right]~,
\end{eqnarray}
and
\begin{equation}
{\partial M\over \partial Q}\biggr\vert _A={Q \over r_+}-\beta{Q \over r_+^3}
{\left(1-{6\over 5}{Q^2\over r_+^2}+{3\over 5}{Q^4\over r_+^4}\right)
\over\left(1-{Q^2\over r_+^2}\right)}~,
\end{equation}
we can obtain the entropy of the black hole given $\Phi_H$ and $T_H$
by Eqs. \ (\ref{3.27}-(\ref{3.28}),
\begin{equation}
S={A\over 4}-{8\pi^2\beta Q^2\over A}+...\label{3.30}
\end{equation}
We observe here that the entropy of the charged black hole is not simply
one quarter of the area, but in general will be a more complicated function
of $A$. The effect of the corrections linear in $\beta$
will then be that of increasing the gravitational entropy by an amount
proportional to $Q^2/A$.
We can also see that the other parameters characterizing the black
hole, such as the charge $Q$ (and the angular momentum if we had considered
rotation (see Ref. \cite{EL93})) enter explicitly in the equation defining
the entropy.

\section{Discussion}

In this paper we have presented a perturbative approach to find solutions
to the classical field equations of a quadratic (in the curvature)
theory of gravitation. Since this theory can be considered a generalization
of general relativity, we start the solutions
from the general relativistic metric and then
add corrections (presumably small) of successive order in the coupling
constants $\alpha$ and $\beta$. It is worth to stress here that our method
does not obtain all the possible solutions to the quadratic theories, but
only those expandable around a metric, solution of the Einstein equations,
in powers of the coupling constants $\alpha$ and $\beta$.
We have thus analyzed the three possible
scenarios of application: Cosmology, Topological Defects and Black Holes,
and obtained the first order corrections to the general relativistic results.
Our method besides allows to make a systematic study of the higher order
corrections by means of symbolic computing programs\cite{cl94}.

The corrected metric (\ref{3.6}) may be of relevance to study the
evolution of gauge strings in the early universe for both the structure
formation scenario where the $r^{-2}$ term in $B(r)$ could play an
important role and for the collision simulations where the short range
contributions may change the predictions of a string network. A detailed
study of these effects might provide a link between observation and the
$\alpha$ and $\beta$ parameters.

The charged black hole metric in quadratic theories gets only modified by
the $\beta$ coupling. The Reissner-Nordstr\"om metric is no longer a solution
to the problem and one must study a different solution (Eq.\ (\ref{3.22})).
The thermodynamics of this charged quadratic black holes is different
from that of a Reissner - Nordstr\"om black hole.
The Bekenstein-Hawking temperature as well as the other thermodynamical
parameters acquire corrections with respect to its general
relativistic values (in Ref. \cite{EL93} they coincide with those of general
relativity) as one would expect for a
different theory of gravitation (see Ref. \cite{MS93} and Ref. \cite{T88}
for a discussion in the context of string and Kaluza-klein theories
respectively).

It is also interesting to remark that for the black hole solution, the relation
between entropy and area is no longer the simple $S={1\over4}A$, but that
given by Eq.\ (\ref{3.30}). we stress that this modification of the
entropy - area
relation have been obtained by considering a four dimensional quadratic
gravitational theory minimally coupled to the electromagnetic field.
We also expect that in an exact black hole solution this simple equation will
break down leaving its place to a more fundamental relation \cite{V93,W93}.

\acknowledgments
This work was partially supported by the Directorate General for
Science, Research and Development of the Commission of the European
Communities. C.O.L was also supported by the Alexander von Humboldt
Foundation and by the Direcci\'on General de
Investigaci\'on Cient\'\i fica y T\'ecnica of the Ministerio de Educaci\'on
y Ciencia de Espa\~na. M.C. holds an scholarship from the Deutscher
Akademischer Austauschdienst.

\end{document}